\documentclass[sigconf,authorversion]{acmart}
\acmConference[ICSE 2024]{46th International Conference on Software Engineering}{April 2024}{Lisbon, Portugal}
\AtBeginDocument{%
  \providecommand\BibTeX{{%
    \normalfont B\kern-0.5em{\scshape i\kern-0.25em b}\kern-0.8em\TeX}}}

\usepackage{enumitem}
\usepackage{tikz} 
\usetikzlibrary{positioning}

\usepackage[]{mdframed}

\begin{document}

\title{Toward an Android Static Analysis Approach for Data Protection}

\author{Mugdha Khedkar}
\affiliation{%
  \institution{\textit{Heinz Nixdorf Institute \\ Paderborn University}}
  \city{Paderborn}
  \country{Germany}
}
\email{mugdha.khedkar@upb.de}

\author{Eric Bodden}
\affiliation{%
  \institution{\textit{Heinz Nixdorf Institute \\ Paderborn University and Fraunhofer IEM}}
  \city{Paderborn}
  \country{Germany}
}
\email{eric.bodden@upb.de}


\begin{abstract}
  Android applications collecting data from users must protect it according to the current legal frameworks. Such data protection has become even more important since the European Union rolled out the General Data Protection Regulation (GDPR). Since app developers are not legal experts, they find it difficult to write privacy-aware source code. Moreover, they have limited tool support to reason about data protection throughout their app development process. 
  
  This paper motivates the need for a static analysis approach to diagnose and explain data protection in Android apps. The analysis will recognize personal data sources in the source code, and aims to further examine the data flow originating from these sources. App developers can then address key questions about data manipulation, derived data, and the presence of technical measures. 
  
  Despite challenges, we explore to what extent one can realize this analysis through static taint analysis, a common method for identifying security vulnerabilities. This is a first step towards designing a tool-based approach that aids app developers and assessors in ensuring data protection in Android apps, based on automated static program analysis.
\end{abstract}



\maketitle

\section{Introduction}
We use several Android applications in our daily life, many of which collect data from us. All Android apps which collect data from users residing in the European Union must comply with the General Data Protection Regulation~\cite{gdpr}, which 
 defines personal data as \textit{``any information relating to an identified or identifiable natural person, a data subject"}. 
 The GDPR imposes several obligations on the access, storage and processing of personal data.

The growing demand for privacy by design~\cite{pbd}, both by end users and by GDPR necessitates that app developers use state-of-the-art technical measures to protect their users’ privacy. The legal description of GDPR is very complex and lengthy and hence it can be difficult for app developers to understand. Since they lack legal expertise, app developers may be left wondering which technical measures need to be taken for which categories of user input data. 

Google Play recently launched the data safety section~\cite{data}, shifting the responsibility of privacy-related reporting to app developers. They
must complete the data safety form, detailing how apps collect, share, and secure users’ data. 
A recent study by Mozilla~\cite{mozilla} has revealed discrepancies between the information reported in data safety sections and privacy policies of Android apps. Accurately disclosing privacy-relevant information 
requires manual effort from app developers. If one can reduce this manual effort, the entire process is likely to become much simpler and more accurate. Thus, tools that bridge the legal and technical aspects of data protection show great potential. 

In this work, we discuss how \emph{static analysis} can be used to design tools for ensuring data protection. Static analysis inspects the source code without executing it, and covers all of the app's \emph{possible} execution paths. Thus the use of static analysis has the potential to eventually yield legally useful guarantees and can be effectively used to aid app developers in writing privacy-aware code. 

Past work has used static \emph{taint analysis} to detect privacy and security violations~\cite{flowdroid,mudflow}. 
For an Android app, taint analysis tracks private data (device identifiers, phone numbers) originating from predefined sources. If such private data reaches predetermined public sinks (database, SMS, internet), it is said to cause a privacy leak. While bare taint analysis is adequate to \emph{detect the presence} of privacy leaks, we will argue in this work that additional analysis support is needed to assess apps for GDPR compliance. This is because developers and assessors must answer questions such as how the app manipulates personal data, which information it derives from that data, and how that information is protected. Taint analysis can support this task but cannot answer these questions on its own.

For GDPR compliance assessment, a static analysis must not only trace personal data flow but also understand how the data is manipulated. 
The idea of tracking data flow from predefined sources is similar to taint analysis. Yet, an important difference to taint analysis is that in the envisioned analysis one needs to explore \emph{all} paths originating at the sources of personal data, and needs to understand the processing activities along those paths. Without a predefined list of sinks, the analysis requires thorough examination of all code actively handling personal data. 



\begin{figure*}
\begin{center}
\includegraphics[width=0.9\textwidth]{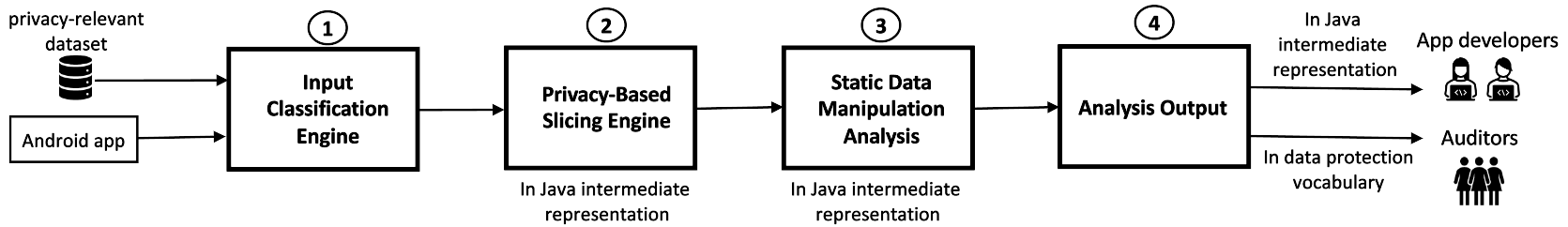}
\caption{Workflow of the envisioned static analysis}
\label{fig:workflow}
\end{center}
\end{figure*}

\section{Workflow and Challenges}
\label{section2}

This section discusses the workflow of the envisioned static analysis (Figure~\ref{fig:workflow}) and the challenges in its design and implementation.
\begin{enumerate}[label=(\arabic*),wide, labelwidth=!, labelindent=0pt]

\item \textit{\textbf{Personal Data Sources.}} To perform any static analysis on potential personal data, one requires a reliable mechanism to detect and label the sources of personal data that an Android app processes. Some of this \textit{system-centric private data}~\cite{uipicker} is provided by the OS, e.g., through system calls like \textit{getLastKnownLocation()}. However, one also needs to label the data that the user provides as input to an Android app, called \textit{user-provided private data}. Different personal data identifiers entail different levels of risks and hence require different levels of data protection. Data identifiers that directly identify a user (eg. passport no., SSN) require other protection than indirect data identifiers (eg. latitude, pincode). Although GDPR considers both direct and indirect identifiers as personal data, separating the two might give more accurate warnings to the app developer. 

The initial step involves implementing an \textit{\textbf{input classification engine}}. This engine labels the sources according to the category of personal data, 
 thereby highlighting the sources that need more critical tracking. We have implemented an engine that filters UI fields from Android apps. Furthermore, we have extended the static code feature extraction tool SootFX~\cite{sootfx} to extract system API calls from the static call graph. Both these components require a \textit{privacy-relevant dataset} which assists in labeling the UI fields and system API calls as personal data sources. To construct this dataset, we collected 350 apps from varying domains. We filtered UI fields of these apps and used SootFX to extract potential personal data sources. Manual labeling resulted in the compilation of the \textit{privacy-relevant dataset}. 
\item \textit{\textbf{Data Disguise.}} GDPR requires app developers to protect personal data by disguising it using a \textit{pseudonymization} function. Pseudonymization replaces parts of personal data with unique, non-identifying \textit{pseudonyms}. Personal data can then no longer be attributed to a specific data subject without the use of pseudonyms~\cite{enisa2}. Pseudonymizing personal data allows organizations to use the data for their purposes while reducing the risk of a privacy breach~\cite{enisa}.

GDPR applies to pseudonymized data but not to anonymized data where the user is unidentifiable. Google also excludes anonymized data from its data safety form. Due to the lack of consensus on the effectiveness of common anonymization techniques~\cite{anonymize}, a Mozilla study~\cite{mozilla} questions Google's approach. True anonymity of data may rely on factors beyond the analysis's scope, leading us to treat anonymized data similarly to pseudonymized data in this work.

To investigate how personal data flows through the source code, one needs to study and record the extent to which personal data is disguised. 
Highlighting pseudonymization functions in the source code will aid app developers in assessing the use of technical measures to protect users' personal data. This will help with source code privacy audits, and answer crucial data protection questions:
\begin{itemize}
\item Is personal data pseudonymized along all paths?
\item Is personal data shared before being pseudonymized?
\end{itemize}

According to GDPR, robust pseudonymization involves creating pseudonymns that cannot be easily \emph{re-identified} and \emph{reproduced} by third parties~\cite{enisa2}. 
Hashing, considered a weak technique, can be reversed with a limited dictionary. Grading pseudonymization functions 
based on these criteria can help app developers identify areas requiring additional data protection measures. 

We use Jicer~\cite{jicer} to understand how personal data flows through the source code. Jicer is a static program slicer that works with an intermediate representation of Java code. 
It takes as input an APK and statically constructs an app dependence graph (ADG), which preserves control and data dependencies in the source code~\cite{slicingSDG}. We are extending Jicer to construct a \textit{\textbf{privacy-based slicing engine}}, which automatically labels \emph{pseudonymization methods} in this graph, and slices this graph from the labeled personal data sources.
\vspace{0.25cm}


\item \textit{\textbf{Data Processing.}} Data protection requires the knowledge of how or where personal data is processed. 
Categorizing methods as privacy-relevant (e.g., analytics, advertising, authorization) is one step, achieved by matching the source code with a dataset of relevant third-party libraries and APIs. However, this is insufficient. After identifying these methods, a thorough examination of how personal data is manipulated around them is necessary. This includes analyzing both third-party methods and those defined within the app, recognizing data processing operations like string or numeric manipulation. 
Generally, data manipulation may include data generation, derivation, retention, 
accumulation, replication or sharing. Labeling these methods is challenging due to the diverse operations possible on personal data. 
Such labeling will, however, aid app developers by answering vital data protection questions:
\begin{itemize}
\item Are multiple indirect identifiers processed in combination?
\item Is data derived from personal data shared with third parties?
\end{itemize}

To address this challenge, we propose a \textit{\textbf{static data manipulation analysis}} which will run on privacy-relevant program slices and examine how labeled personal data is manipulated through the source code. The analysis will inform the developer/assessor about the nature of the identified data manipulation (e.g. derivation, sharing, etc.), thereby easing the task of privacy assessment.
\vspace{0.25cm}

\item \textit{\textbf{Analysis Output.}} 
The envisioned analysis will 
target both legal and technical experts. It will do so by visualizing privacy-relevant program slices that represent how personal data flows through and is processed in the source code. The visualization will support abstract views that express source code components in terms of legal aspects of GDPR using a data protection vocabulary~\cite{dpv}. Such views can effectively support a risk analysis of the program slices and highlight parts of the application code that process personal data. A possible risk, for instance, would be if program parts achieve such processing through code from third parties that are untrusted. More detailed views will allow developers to understand the data processing directly in their program code.
\end{enumerate}

\section{Case Study}
\label{section3}

We studied the privacy policy and data safety section of two free planetarium apps available on Google Play Store: Stellarium\footnote{\url{https://stellarium-labs.com/stellarium-mobile-plus/}} and SkyMap\footnote{\url{https://play.google.com/store/apps/details?id=com.google.android.stardroid}}. Their privacy policy and data safety information both claim to \emph{not} collect \emph{any} personal data from their users. For both these apps, while the data safety section claims that data is encrypted in transit, the privacy policy does not mention encryption at all. SkyMap's privacy policy claims that the app shares all \emph{anonymous data} with Google Analytics. On the other hand, its data safety section claims to not share \emph{any} data with any third parties. So apparently neither documentation provides a usable ground truth.

Our observations are summarized below:
\begin{itemize}
\item The \textit{input classification engine} confirmed that these apps indeed do \emph{not} collect data from their user interfaces. 
However, for both these apps it detected \emph{system-centric personal data} sources, which collect location and device data which can indirectly identify the user. For SkyMap, we also found \emph{Google account methods} collecting email address and other account details from the users.
\item The \textit{privacy-based slicing engine} detected pseudonymization methods from the ADG of both these apps. While  SkyMap uses \emph{<java.security.MessageDigest>} for pseudonymization, we observed the presence of \emph{<javax.crypto.Cipher>} in Stellarium's source code.
\item SkyMap's app dependence graph labeled by the \textit{privacy-based slicing engine} revealed the usage of \emph{Firebase Analytics}, among other Google APIs for analytics and advertisements.
\end{itemize}

These early results are evidence that even apps that claim to not collect any personal data are processing some user data, but also hint towards the presence of data protection measures. We further aim to slice the ADG to observe how these sources of personal data flow through the source code. A data manipulation analysis that runs on these privacy-relevant program slices will help us answer the following questions with respect to the above two apps:
\begin{itemize}
    \item Is the personal data collected by these apps pseudonymized?
    \item How are the user's Google account credentials processed in SkyMap? Why were they required in the first place?
    \item SkyMap uses message digest. However, commonly used hash functions such as MD5 and SHA-1~\cite{crypto} with known vulnerabilities with respect to the probability of finding collisions should be avoided~\cite{enisa2}. This raises the question: \emph{Are these apps really using \textbf{robust} pseudonymization functions?}
    \item Which data does SkyMap share with Google Analytics? 
    \item SkyMap's privacy policy claims to share \emph{anonymous data} with Analytics. How is data anonymized in the source code?
\end{itemize}
Answering such questions will enhance privacy transparency and assist app developers in prioritizing data protection along the whole app development process.
\section{Related Work}
\label{section4}


Existing tools~\cite{supor, uipicker} can identify sensitive user-provided data in Android apps but do not categorize it as personal data. Since these tools are proprietary, they cannot be used in our work. 

Most static taint analysis tools~\cite{flowdroid,iccta,mudflow} use predefined lists of SOURCE and SINK methods as a starting point for their analysis. 
In the next years, machine-learning approaches were proposed to classify and categorize methods as sources and sinks~\cite{susi,swan,ase19swanAssist,codoc}. 
Kober et al.~\cite{sensitive_data} provide a sound definition of sensitive data derived from the definition of personal data of several legal frameworks (including GDPR). They publicly provide a list of sensitive sources from the Android framework. Our idea differs from all the above-mentioned techniques in two ways. First, we will categorize data as personal data with respect to GDPR. Secondly, we are interested in both user-provided and system-centric private data.

Feiyang Tang and Bjarte M. Østvold~\cite{privflow} recently introduced an automatic software analysis technique that characterizes the flow of privacy-related data, presenting results as a graph comprehensible to software developers and auditors. However, their analysis is aimed at Data Protection Impact Management~\cite{dpia} and answers only limited questions. PTPDroid~\cite{PTPDroid} uses the taint analysis tool FlowDroid~\cite{flowdroid} to study the reachability of third party libraries and compares it with the privacy policy section of "data sharing with third parties". ATPChecker~\cite{atpchecker} uses FlowDroid to automatically identify whether the usage of in-app third-party libraries complies with privacy-related regulations. All these tools~\cite{privflow,PTPDroid,atpchecker} use predefined lists of SOURCE and SINK methods, and do not categorize data as personal data. Moreover, PTPDroid and ATPChecker focus on data sharing and lack examination of in-app data manipulation. We further plan to highlight where personal data is disguised and how in-app methods manipulate personal data.

\section{Future Plans and Conclusion}
\label{section6}

We will conduct an empirical evaluation of our \textbf{input classification engine}, examining privacy-relevant data collection across several domains on Android apps.

Simultaneously, we are enhancing the visualization of the \textbf{privacy-based slicing engine} for better understanding of app developers. User studies with Android developers will ensure the usability and comprehensibility of our engine. We will further design an abstract visualization that can be understood by legal experts. We will extend these visualizations to present the final \textbf{analysis results} once the analysis has been completely implemented.

Once the visualization is usable, we will examine the privacy-relevant program slices of Android apps to derive tool automation that can statically detect, classify, and visualize data manipulation, resulting in our \textbf{static data manipulation analysis}. 


The envisioned analysis will enable developers and assessors of Android apps to rapidly understand how and where an app processes personal data. This assistance is valuable for manually validating program behavior against the app's privacy policy.

\bibliographystyle{ACM-Reference-Format}
\bibliography{sample-base}
\end{document}